\documentclass{article}
\usepackage{spconf,amsmath,graphicx, booktabs, array, subfig}
\usepackage{amssymb}
\usepackage{pifont}
\usepackage{xcolor}
\usepackage[colorlinks,allcolors=blue]{hyperref}
\usepackage{verbatim}

\newcommand{\chmark}{\ding{51}}
\newcommand{\crmark}{\ding{55}}
\newcommand{\source}[1]{\texttt{#1}}
\newcommand{\alex}[1]{{\color{blue} A: #1}}
\newcommand{\simon}[1]{{\color{red} S: #1}}

\title{hybrid transformers for music source separation}
\name{Simon Rouard, Francisco Massa, Alexandre D{\'{e}}fossez}
\address{Meta AI}
\begin{document}
%
\maketitle
\begin{abstract}
A natural question arising in Music Source Separation (MSS) is whether long range contextual information is useful,
or whether local acoustic features are sufficient.
In other fields, attention based Transformers~\cite{transformer} have shown their ability to integrate
information over long sequences. In this work, we introduce Hybrid Transformer Demucs (HT Demucs), 
an hybrid temporal/spectral bi-U-Net based on Hybrid Demucs~\cite{defossez2021hybrid}, 
where the innermost layers are replaced by a cross-domain Transformer Encoder, 
using self-attention within one domain, and cross-attention across domains.
While it performs poorly when trained only on MUSDB~\cite{musdb}, we show that it outperforms Hybrid Demucs (trained on the same data) by 0.45 dB of SDR when using 800 extra training songs.
Using sparse attention kernels to extend its receptive field, and per source fine-tuning, 
we achieve state-of-the-art results on MUSDB with extra training data, with 9.20 dB of SDR.

\end{abstract}
\begin{keywords}
Music Source Separation, Transformers
\end{keywords}
\section{Introduction}
\label{sec:intro}

Since the 2015 Signal Separation Evaluation Campaign (SiSEC) \cite{sisec15}, the community of MSS has mostly focused on the task of training supervised models to separate songs into 4 stems: drums, bass, vocals and other (all the other instruments). The reference dataset that is used to benchmark MSS is MUSDB18~\cite{musdb,musdb18-hq} which is made of 150 songs in two versions (HQ and non-HQ). 
Its training set is composed of 87 songs, a relatively small corpus compared with other deep learning based tasks, where Transformer~\cite{transformer} based architectures have seen widespread success and adoption, such as vision~\cite{layerscale,Rombach_2022_CVPR} or natural language tasks~\cite{brown2020language}. Source separation is a task where having a short context or a long context as input both make sense. Conv-Tasnet~\cite{convtasnet} uses about one second of context to perform the separation, using only local acoustic features. On the other hand, Demucs~\cite{demucsv2} can use up to 10 seconds of context, which can help to resolve ambiguities in the input. In the present work, we aim at studying how Transformer architectures can help leverage this context, and what amount of data is required to train them.

We first present in Section~\ref{sec:architecture} a novel architecture, \emph{Hybrid Transformer Demucs} (HT Demucs), which replaces the innermost layers of the original Hybrid Demucs architecture~\cite{defossez2021hybrid} with Transformer layers, applied both in the time and spectral
representation, using self-attention within one domain, and cross-attention across domains.
As Transformers are usually data hungry, we leverage an internal dataset composed of 800 songs on top of the MUSDB dataset, described in Section~\ref{sec:dataset}. 

Our second contribution is to evaluate extensively this new architecture in Section~\ref{sec:results}, with various settings (depth, number of channels, context length, augmentations etc.). We show in particular that it improves over the baseline Hybrid Demucs architecture (retrained on the same data) by 0.35~dB. 

Finally, we experiment with increasing the context duration using sparse kernels based with Locally Sensitive Hashing
to overcome memory issues during training, and fine-tuning procedure, thus achieving a final SDR of 9.20~dB on the test set of MUSDB. 

We release the training code, pre-trained models, and samples on our github~\href{https://github.com/facebookresearch/demucs}{facebookresearch/demucs.}


\section{Related Work}
\label{sec:related}

A traditional split for MSS methods is between spectrogram based and waveform based models.
The former includes models like Open-Unmix~\cite{umx}, a biLSTM with fully connected that
predicts a mask on the input spectrogram or D3Net~\cite{d3net} which uses dilated convolutional blocks with dense connections. More recently, using complex-spectrogram as input and output was favored~\cite{lasaft} as it provides a richer representation and removes the topline given by the Ideal-Ratio-Mask.
The latest spectrogram model, Band-Split RNN~\cite{bsrnn}, combines this idea, along with multiple
dual-path RNNs~\cite{luo2020dual}, each acting in carefully crafted frequency band. 
It currently achieves the state-of-the-art on MUSDB with 8.9 dB.
Waveform based models started with Wave-U-Net~\cite{waveunet}, which served as the basis for Demucs~\cite{demucsv2}, a 
time domain U-Net with a bi-LSTM between the encoder and decoder. Around the same time, Conv-TasNet
showed competitive results~\cite{convtasnet,demucsv2} using residual dilated convolution blocks to predict
a mask over a learnt representation. Finally, a recent trend has been to use both temporal and spectral domains, either through model blending, like KUIELAB-MDX-Net~\cite{kuielab}, or using
a bi-U-Net structure with a shared backbone as Hybrid Demucs~\cite{defossez2021hybrid}. Hybrid Demucs
was the first ranked architecture at the latest MDX MSS Competition~\cite{mdx2021}, although
it is now surpassed by Band-Split RNN.

Using large datasets has been shown to be beneficial to the task of MSS.
Spleeter \cite{spleeter} is a spectrogram masking U-Net architecture trained on 25,000 songs extracts of 30 seconds,
and was at the time of its release, the best model available. Both D3Net and Demucs highly benefited from
using extra training data, while still offering strong performance on MUSDB only.
Band-Split RNN introduced a novel unsupervised augmentation technique requiring only mixes to improve its 
performance by 0.7 dB of SDR.

Transformers have been used for speech source separation with SepFormer~\cite{subakan2021attention},  which is similar to Dual-Path RNN: short range attention layers are interleaved with long range ones.
However, its requires almost 11GB of memory for the forward pass for 5 seconds of audio at 8 kHz, and thus is not 
adequate for studying longer inputs at 44.1 kHz.

\begin{table}
\caption{Comparison with baselines on the test set of MUSDB HQ (methods with a $^*$ are reported on the non HQ version). ``Extra?'' indicates the number of 
extra songs used at train time, $\dagger$ indicates that only mixes are used. ``fine tuned'' indices
per source fine-tuning.}
\label{tab:baselines}
\begin{center}
\resizebox{0.48\textwidth}{!}{
\begin{tabular}{l c r r r r r}
  \toprule
     &&& \multicolumn{4}{c}{Test SDR in dB}\\
     \cmidrule{3-7}
  \textbf{Architecture} & \textbf{Extra?} &
  \source{All} & \source{Drums} &  \source{Bass} &\source{Other} & \source{Vocals}\\
  \midrule
  IRM oracle & N/A & 8.22 & 8.45 & 7.12 & 7.85 & 9.43\\
  \midrule
  KUIELAB-MDX-Net \cite{kuielab} & \crmark & 7.54 &7.33&7.86&5.95& 9.00 \\
  Hybrid Demucs \cite{defossez2021hybrid} & \crmark & 7.64 & 8.12 & 8.43 & 5.65 & 8.35 \\
  Band-Split RNN \cite{bsrnn} & \crmark & \textbf{8.24} & 9.01 & 7.22 & 6.70 & 10.01 \\
  \midrule
  HT Demucs & \crmark & 7.52 & 7.94 & 8.48 & 5.72 & 7.93  \\
  \midrule
  Spleeter$^*$ \cite{spleeter} & $25\mathrm{k}$ & 5.91 & 6.71 & 5.51 & 4.55 & 6.86\\
  D3Net$^*$ \cite{d3net} & 1.5k & 6.68 & 7.36  & 6.20 & 5.37 & 7.80 \\
  Demucs v2$^*$ \cite{demucsv2} & 150 & 6.79 & 7.58 & 7.60 & 4.69 & 7.29  \\
  Hybrid Demucs \cite{defossez2021hybrid} & 800 & 8.34 &9.31&9.13& 6.18& 8.75  \\
  Band-Split RNN \cite{bsrnn} & 1750$^\dagger$ & 8.97 & 10.15 & 8.16 & \textbf{7.08} & \textbf{10.47} \\
  \midrule
  HT Demucs & 150 & 8.49 & 9.51 & 9.76 & 6.13 & 8.56  \\
  HT Demucs & 800 & 8.80 & 10.05 & 9.78 & 6.42 & 8.93  \\
  HT Demucs (fine tuned) & 800 & 9.00 & 10.08 & 10.39 & 6.32 & 9.20  \\
  Sparse HT Demucs (fine tuned) & 800 & \textbf{9.20} & \textbf{10.83} & \textbf{10.47} & 6.41 & 9.37  \\
  \bottomrule
\end{tabular}}
\end{center}
\vspace{-0.1cm}
{\footnotesize }
\end{table}

\section{Architecture}
\label{sec:architecture}

\begin{figure*}[h]
    \centering
    \subfloat[\centering Transformer Encoder Layer]{{\includegraphics[width=2.6cm]{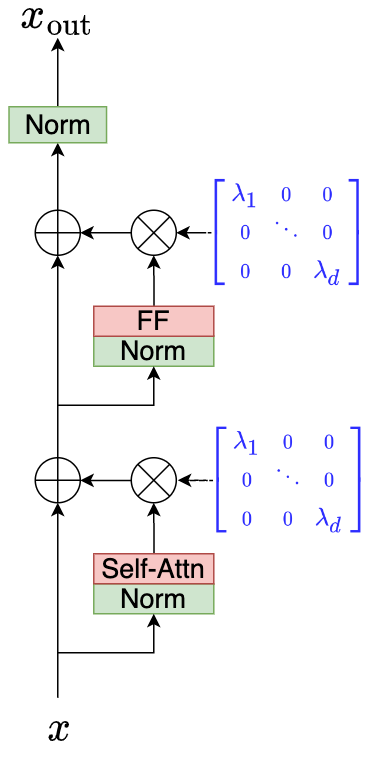} }}%
    \qquad
    \subfloat[\centering Cross-domain Transformer Encoder of depth 5]{{\includegraphics[width=5cm]{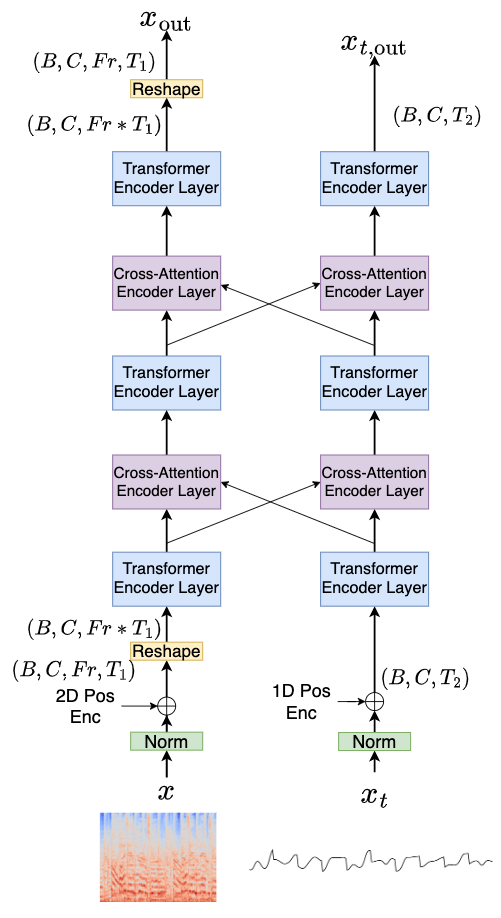} }}%
    \qquad
    \subfloat[\centering Hybrid Transformer Demucs]{{\includegraphics[width=8.5cm]{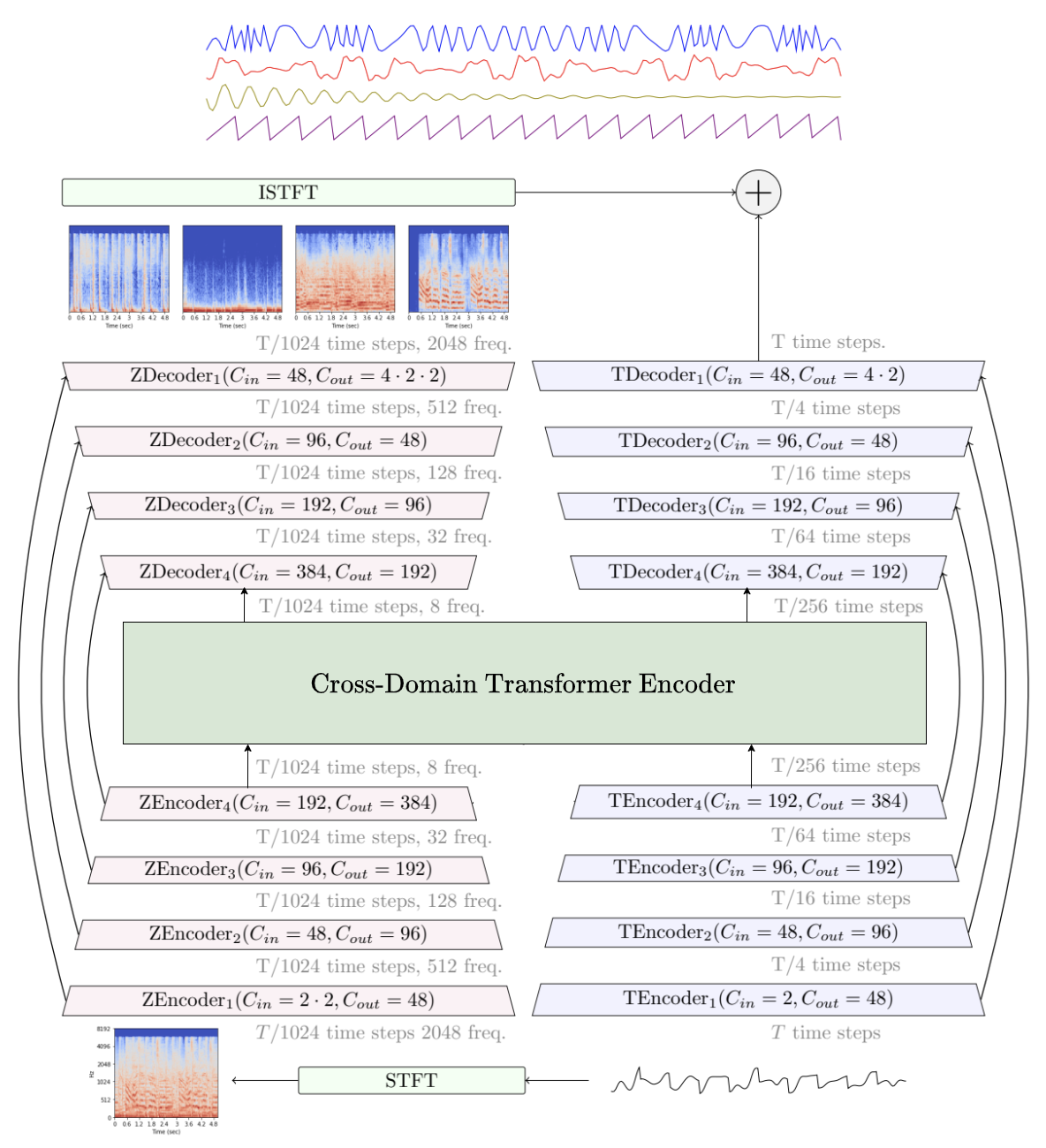} }}%
    \caption{Details of the Hybrid Transformer Demucs architecture. (a): the Transformer Encoder layer
        with self-attention and Layer Scale \cite{layerscale}. (b): The Cross-domain Transformer Encoder treats spectral and temporal signals with interleaved Transformer Encoder layers and cross-attention Encoder layers. (c): Hybrid Transformer Demucs keeps the outermost 4 encoder and decoder layers of Hybrid Demucs with the addition of a cross-domain Transformer Encoder between them.}
    \label{fig:architecture}%
\end{figure*}

We introduce the Hybrid Transformer Demucs model, based on Hybrid Demucs~\cite{defossez2021hybrid}.
The original Hybrid Demucs model is made of two U-Nets, one in the time domain (with temporal convolutions) and
one in the spectrogram domain (with convolutions over the frequency axis). 
Each U-Net is made of 5 encoder layers, and 5 decoder layers. After the 5-th encoder layer, both representation
have the same shape, and they are summed before going into a shared 6-th layer. Similarly, the first decoder layer is shared, and
its output is sent both the temporal and spectral branch. The output of the spectral branch
is transformed to a waveform using the iSTFT, before being summed with the output of the temporal branch, giving the 
actual prediction of the model.

Hybrid Transformer Demucs keeps the outermost 4 layers as is from the original architecture, and replaces the 2 innermost layers in the encoder and the decoder, including local attention and bi-LSTM, with
a cross-domain Transformer Encoder.
It treats in parallel the 2D signal from the spectral branch and the 1D signal from the waveform branch. 
Unlike the original Hybrid Demucs which required careful tuning of the model parameters (STFT window and hop length, stride, paddding, etc.) to align the time and spectral representation, the cross-domain
Transformer Encoder can work with heterogeneous data shape, making it a more flexible architecture.

The architecture of Hybrid Transformer Demucs is depicted on Fig.\ref{fig:architecture}. On the left, we show a single self-attention Encoder layer of the Transformer~\cite{transformer} with normalizations before the Self-Attention and Feed-Forward operations, it is combined with Layer Scale \cite{layerscale} initialized to $\epsilon{=}10^{-4}$ in order to stabilize the training. The two first normalizations are layer normalizations (each token is independently normalized) and the third one is a time layer normalization (all the tokens are normalized together). 
The input/output dimension of the Transformer is $384$, and linear layers are used to convert to the internal
dimension of the Transformer when required.
The attention mechanism has 8 heads and the hidden state size of the feed forward network is equal to 4 times the dimension of the transformer. The cross-attention Encoder layer is the same but using cross-attention with the other domain representation. In the middle, a cross-domain Transformer Encoder of depth 5 is depicted. It is the interleaving of self-attention Encoder layers and cross-attention Encoder layers in the spectral and waveform domain. 1D \cite{transformer} and 2D \cite{2Dpe} sinusoidal encodings are added to the scaled inputs and reshaping is applied to the spectral representation in order to treat it as a sequence. On Fig.~\ref{fig:architecture} (c), we give
a representation of the entire architecture, along with the double U-Net encoder/decoder structure.

Memory consumption and the speed of attention quickly deteriorates with an increase of the sequence lengths.
To further scale, we leverage sparse attention kernels introduced in the \texttt{xformer} package~\cite{xFormers2021},
along with a Locally Sensitive Hashing (LSH) scheme to determine dynamically the sparsity pattern. We use a sparsity level
of 90\% (defined as the proportion of elements removed in the softmax), which is determined by performing 32 rounds of LSH with 4 buckets each. We select the elements that match at least $k$ times over all 32 rounds of LSH, with $k$ such that the sparsity level is 90\%.
We refer to this variant as Sparse HT Demucs.

\section{Dataset}
\label{sec:dataset}
We curated an internal dataset composed of 3500 songs with the stems from 200 artists with diverse music genres. Each stem is assigned to one of the 4 sources according to the name given by the music producer (for instance "vocals2", "fx", "sub" etc...). This labeling is noisy because these names are subjective and sometime ambiguous. 
For 150 of those tracks, we manually verified that the automated 
labeling was correct, and discarded ambiguous stems. 
We trained a first Hybrid Demucs model on MUSDB and those 150 tracks.
We preprocess the dataset according to several rules.
First, we keep only the stems for which all four sources are non silent at least 30\% of the time. For each 1 second segment, we define it as silent if its volume is less than -40dB. 
Second, for a song $x$ of our dataset, noting $x_i$, with $i \in \{\mathrm{drums, bass, other, vocals} \}$, each stem and $f$ the Hybrid Demucs model previously mentioned, we define $y_{i,j}{=}f(x_i)_j$ i.e. the output $j$ when separating the stem $i$. Theoretically, if all the stems were perfectly labeled and if $f$ were a perfect source separation model we would have $y_{i, j} = x_i \delta_{i, j}$ with $\delta_{i, j}$ being the Kronecker delta.
For a waveform $z$, let us define the volume in dB measured over 1 second segments:
\begin{equation}
    V(z) = 10\cdot\log_{10}\left(\mathrm{AveragePool}(z^2, 1\text{ sec})\right).
\end{equation}
For each pair of sources $i,j$, we take the segments of 1 second where the stem is present (regarding the first criteria) and define $P_{i,j}$ as the proportion of these segments where $V(y_{i,j}) - V(x_i) > -10\,\mathrm{dB}$.
We obtain a square matrix $P\in [0, 1]^{4 \times 4}$, and we notice that in perfect condition, we should have $P = \mathrm{Id}$. We thus keep only the songs for which for all sources $i$, $P_{i, i} > 70\%$, and pairs
of sources $i \neq j$, $P_{i, j} < 30\%$. This procedure selects 800 songs.

\section{Experiments and Results}
\label{sec:results}
\subsection{Experimental Setup}
All our experiments are done on 8 Nvidia V100 GPUs with 32GB of memory, using fp32 precision. We use the L1 loss on the waveforms, optimized with Adam \cite{adam} without weight decay, a learning rate of $3 \cdot 10^{-4}$, $\beta_1 = 0.9$, $\beta_2 = 0.999$ and a batch size of 32 unless stated otherwise. We train for 1200 epochs of 800 batches each over the MUSDB18-HQ dataset, completed with
the 800 curated songs dataset presented in Section~\ref{sec:dataset}, sampled at 44.1kHz and stereophonic. We use exponential moving average
as described in~\cite{defossez2021hybrid}, and select the best model over the valid set, composed of the valid set
of MUSDB18-HQ and another 8 songs.
We use the same data augmentation as described in~\cite{defossez2021hybrid}, including repitching/tempo stretch
and remixing of the stems within one batch.

Using one model per source can be beneficial~\cite{bsrnn}, although its adds overhead both at train time
and evaluation. In order to limit the impact at train time, we propose a  procedure
where one copy of the multi-target model is fine-tuned on a single target task for 50 epochs, with a learning rate of $10^{-4}$, no remixing, repitching, nor rescaling applied as data augmentation.
Having noticed some instability towards the end of the training of the main model, we use for the fine tuning a gradient clipping (maximum L2 norm of $5.$ for the gradient), and a weight decay of 0.05. 

At test time, we split the audio into chunks having the same duration as the one used for training, with an overlap of 25\% and a linear transition from one chunk to the next.
We report is the Signal-to-Distortion-Ratio (SDR) as defined by the SiSEC18 \cite{sisec18} which is the median across the median SDR over all 1 second chunks in each song. On Tab.~\ref{tab:preliminary} we also report the Real Time Factor (RTF)
computed on a single core of an Intel Xeon CPU at 2.20GHz. It is defined as the time to process some fixed audio input (we use 40 seconds of gaussian noise) divided by the input duration.

\subsection{Comparison with the baselines}

On Tab.~\ref{tab:baselines}, we compare to several state-of-the-art baselines. For reference, we also provide
baselines that are trained without any extra training data. Comparing to the original Hybrid Demucs architecture,
we notice that the improved sequence modeling capabilities of transformers increased the SDR by 0.45 dB for the simple version, and up to almost 0.9 dB when using the sparse variant of our model along with fine tuning.
The most competitive baseline is Band-Split RNN~\cite{bsrnn}, which achieves a better SDR on both the other and vocals sources, despite using only MUSDB18HQ as a supervised training set, and using 1750 unsupervised tracks as extra training data.

\subsection{Impact of the architecture hyper-parameters}
\label{sec:hyperparams}

We first study the influence of three architectural hyper-parameters in Tab.~\ref{tab:preliminary}: the duration in seconds of the excerpts used to train the model, the depth of the transformer encoder and its dimension.
For short training excerpts (3.4 seconds), we notice that augmenting the depth from 5 to 7 increases the test SDR and that augmenting the transformer dimension from 384 to 512 lowers it slightly by 0.05 dB. With longer segments (7.8 seconds), we observe an increase of almost 0.6 dB when using a depth of 5 and 384 dimensions. Given this observation, we also tried to increase the duration or the depth but this led to Out Of Memory (OOM) errors. Finally, augmenting the dimension to 512 when training over 7.8 seconds led to an improvement of 0.1 dB. 


\subsection{Impact of the data augmentation}
\label{sec:data_augment}

On Tab.~\ref{tab:mixing}, we study the impact of disabling some of the data augmentation, 
as we were hoping that using more training data would reduce the need for such augmentations. However,
we observe a constant deterioration of the final SDR as we disable those.
While the repitching augmentation has a limited impact, the remixing augmentation
remain highly important to train our model, with a loss of 0.7 dB without it.

\subsection{Impact of using sparse kernels and fine tuning}
\label{sec:finetune}

We test the sparse kernels described in Section~\ref{sec:architecture} to increase the depth to 7 and
the train segment duration to 12.2 seconds, with a dimension of 512. This simple change yields an extra 0.14 dB of SDR (8.94 dB). The fine-tuning per source improves the SDR by 0.25 dB, to 9.20 dB, despite requiring only 50 epochs to train. We tried further extending the receptive field
of the Transformer Encoder to 15 seconds during the fine tuning stage by reducing the batch size, 
however, this led to the same SDR of 9.20 dB, although training
from scratch with such a context might lead to a different result.

\section*{Conclusion}

We introduced Hybrid Transformer Demucs, a Transformer based variant of Hybrid Demucs that replaces the innermost convolutional layers by a Cross-domain Transformer Encoder, using self-attention and cross-attention to process spectral and temporal informations. This architecture benefits from our large training dataset and outperforms Hybrid Demucs by 0.45 dB. Thanks to sparse attention techniques, we scaled our model to an input length up to 12.2 seconds during training which led to a supplementary gain of 0.4 dB. 
 Finally, we could explore splitting the spectrogram into subbands in order to process them differently as it is done in \cite{bsrnn}.

\begin{table}
\caption{Impact of segment duration, transformer depth and transformer dimension. OOM means Out of Memory. Results are commented in Sec.~\ref{sec:hyperparams}.}
\label{tab:preliminary}
\begin{center}
\resizebox{0.46\textwidth}{!}{
\begin{tabular}{r c c r c r}
  \toprule
  \textbf{dur. (sec)} & \textbf{depth} & \textbf{dim.} &
 \textbf{nb. param.} & \textbf{RTF (cpu)} &  \source{SDR (All)} \\
  \midrule
  3.4 & 5 & 384 &  26.9M & 1.02 & 8.17 \\ 
  3.4 & 7 & 384 &  34.0M & 1.23 & 8.26 \\ 
  3.4 & 5 & 512 &  41.4M & 1.30 & 8.12\\ 
  \midrule
  7.8 & 5 & 384 &  26.9M & 1.49 & 8.70\\ 
  7.8 & 7 & 384 &  34.0M & 1.68 & OOM\\ 
  7.8 & 5 & 512 &  41.4M & 1.77 & \textbf{8.80} \\ 
  \midrule
  12.2 & 5 & 384 &  26.9M & 2.04 & OOM\\ 
  \bottomrule
\end{tabular}}
\end{center}
\end{table}

\begin{table}
\caption{Impact of data augmentation. The model has a depth of 5, and a dimension of 384. See Sec.~\ref{sec:data_augment} for details.}
\label{tab:mixing}
\begin{center}
\resizebox{0.42\textwidth}{!}{
\begin{tabular}{c c c c c}
  \toprule
  \textbf{dur. (sec)} & \textbf{depth} & \textbf{remixing} & \textbf{repitching} &
  \source{SDR (All)} \\
  \midrule
  7.8 & 5 & \chmark & \chmark & 8.70 \\  
  7.8  &5 & \chmark & \crmark & 8.65 \\
  7.8  & 5& \crmark & \chmark & 8.00 \\

  \bottomrule
\end{tabular}}
\end{center}
\end{table}

\clearpage
\bibliographystyle{IEEEbib}
\bibliography{refs}

\begin{thebibliography}{10}

\bibitem{transformer}
Ashish Vaswani, Noam Shazeer, Niki Parmar, Jakob Uszkoreit, Llion Jones,
  Aidan~N. Gomez, Lukasz Kaiser, and Illia Polosukhin,
\newblock ``Attention is all you need,''
\newblock {\em CoRR}, vol. abs/1706.03762, 2017.

\bibitem{defossez2021hybrid}
Alexandre D{\'e}fossez,
\newblock ``Hybrid spectrogram and waveform source separation,''
\newblock in {\em Proceedings of the ISMIR 2021 Workshop on Music Source
  Separation}, 2021.

\bibitem{musdb}
Zafar Rafii, Antoine Liutkus, Fabian-Robert Stöter, Stylianos~Ioannis
  Mimilakis, and Rachel Bittner,
\newblock ``The musdb18 corpus for music separation,'' 2017.

\bibitem{sisec15}
Nobutaka Ono, Zafar Rafii, Daichi Kitamura, Nobutaka Ito, and Antoine Liutkus,
\newblock ``{The 2015 Signal Separation Evaluation Campaign},''
\newblock in {\em {International Conference on Latent Variable Analysis and
  Signal Separation (LVA/ICA)}}, Aug. 2015.

\bibitem{musdb18-hq}
Zafar Rafii, Antoine Liutkus, Fabian-Robert Stöter, Stylianos~Ioannis
  Mimilakis, and Rachel Bittner,
\newblock ``Musdb18-hq - an uncompressed version of musdb18,'' Aug. 2019.

\bibitem{layerscale}
Hugo Touvron, Matthieu Cord, Alexandre Sablayrolles, Gabriel Synnaeve, and
  Herv{\'e} J{\'e}gou,
\newblock ``Going deeper with image transformers,''
\newblock in {\em Proceedings of the IEEE/CVF International Conference on
  Computer Vision}, 2021.

\bibitem{Rombach_2022_CVPR}
Robin Rombach, Andreas Blattmann, Dominik Lorenz, Patrick Esser, and Bj\"orn
  Ommer,
\newblock ``High-resolution image synthesis with latent diffusion models,''
\newblock in {\em Proceedings of the IEEE/CVF Conference on Computer Vision and
  Pattern Recognition (CVPR)}, June 2022, pp. 10684--10695.

\bibitem{brown2020language}
Tom B.~Brown et~al.,
\newblock ``Language models are few-shot learners,''
\newblock 2020.

\bibitem{convtasnet}
Yi~Luo and Nima Mesgarani,
\newblock ``Conv-tasnet: Surpassing ideal time--frequency magnitude masking for
  speech separation,''
\newblock {\em IEEE/ACM Transactions on Audio, Speech, and Language
  Processing}, 2019.

\bibitem{demucsv2}
Alexandre Défossez, Nicolas Usunier, Léon Bottou, and Francis Bach,
\newblock ``Music source separation in the waveform domain,'' 2019.

\bibitem{umx}
F.-R. St\"oter, S.~Uhlich, A.~Liutkus, and Y.~Mitsufuji,
\newblock ``Open-unmix - a reference implementation for music source
  separation,''
\newblock {\em Journal of Open Source Software}, 2019.

\bibitem{d3net}
Naoya Takahashi and Yuki Mitsufuji,
\newblock ``D3net: Densely connected multidilated densenet for music source
  separation,'' 2020.

\bibitem{lasaft}
Woosung Choi, Minseok Kim, Jaehwa Chung, and Soonyoung Jung,
\newblock ``Lasaft: Latent source attentive frequency transformation for
  conditioned source separation,''
\newblock in {\em IEEE International Conference on Acoustics, Speech and Signal
  Processing (ICASSP)}, 2021.

\bibitem{bsrnn}
Yi~Luo and Jianwei Yu,
\newblock ``Music source separation with band-split rnn,'' 2022.

\bibitem{luo2020dual}
Yi~Luo, Zhuo Chen, and Takuya Yoshioka,
\newblock ``Dual-path rnn: efficient long sequence modeling for time-domain
  single-channel speech separation,''
\newblock in {\em ICASSP 2020-2020 IEEE International Conference on Acoustics,
  Speech and Signal Processing (ICASSP)}. IEEE, 2020, pp. 46--50.

\bibitem{waveunet}
Daniel Stoller, Sebastian Ewert, and Simon Dixon,
\newblock ``Wave-u-net: A multi-scale neural network for end-to-end audio
  source separation,''
\newblock {\em arXiv preprint arXiv:1806.03185}, 2018.

\bibitem{kuielab}
Minseok Kim, Woosung Choi, Jaehwa Chung, Daewon Lee, and Soonyoung Jung,
\newblock ``Kuielab-mdx-net: A two-stream neural network for music demixing,''
  2021.

\bibitem{mdx2021}
Yuki Mitsufuji, Giorgio Fabbro, Stefan Uhlich, Fabian-Robert Stöter, Alexandre
  D{\'{e}}fossez, Minseok Kim, Woosung Choi, Chin-Yun Yu, and Kin-Wai Cheuk,
\newblock ``Music demixing challenge 2021,''
\newblock {\em Frontiers in Signal Processing}, vol. 1, jan 2022.

\bibitem{spleeter}
Romain Hennequin, Anis Khlif, Felix Voituret, and Manuel Moussallam,
\newblock ``Spleeter: a fast and efficient music source separation tool with
  pre-trained models,''
\newblock {\em Journal of Open Source Software}, 2020.

\bibitem{subakan2021attention}
Cem Subakan, Mirco Ravanelli, Samuele Cornell, Mirko Bronzi, and Jianyuan
  Zhong,
\newblock ``Attention is all you need in speech separation,''
\newblock in {\em ICASSP 2021-2021 IEEE International Conference on Acoustics,
  Speech and Signal Processing (ICASSP)}. IEEE, 2021, pp. 21--25.

\bibitem{2Dpe}
Zelun Wang and Jyh-Charn Liu,
\newblock ``Translating math formula images to latex sequences using deep
  neural networks with sequence-level training,'' 2019.

\bibitem{xFormers2021}
Benjamin Lefaudeux, Francisco Massa, Diana Liskovich, Wenhan Xiong, Vittorio
  Caggiano, Sean Naren, Min Xu, Jieru Hu, Marta Tintore, and Susan Zhang,
\newblock ``xformers: A modular and hackable transformer modelling library,''
  \url{https://github.com/facebookresearch/xformers}, 2021.

\bibitem{adam}
Diederik~P. Kingma and Jimmy Ba,
\newblock ``Adam: A method for stochastic optimization,'' 2014.

\bibitem{sisec18}
Fabian-Robert Stöter, Antoine Liutkus, and Nobutaka Ito,
\newblock ``The 2018 signal separation evaluation campaign,'' 2018.

\end{thebibliography}

\end{document}